# Large-scale LH$_2$ pipeline infrastructure concept for airports

H. A. Krog[a], Y. Jooss[a], H. Fyhn[a], P. Nekså[a], I. Hjorth[a]

[a] *SINTEF Energy Research, Trondheim, 7465, Norway*
e-mail: halvor.krog@sintef.no

Abstract

Infrastructure and processes for handling of liquid hydrogen (LH$_2$) is needed to enable large-scale decarbonization of aviation with hydrogen aircraft. At large airports, pipeline and hydrant systems will be important for a mature hydrogen-powered air travel market. As the vaporization of LH$_2$ is a challenge in fuel handling, the pipeline infrastructure must be designed and operated such that the fuel is subcooled. Through modelling and simulation of aircraft tanks refuelling by a pipeline infrastructure concept, it is found that continuous recycling of LH$_2$ within the system is needed to maintain subcooling, and the pump operation is important for preventing flashing. With the proposed concept, some hydrogen vapor is formed in the aircraft tank, but the vapor can be utilised by hydrogen-powered ground support equipment.

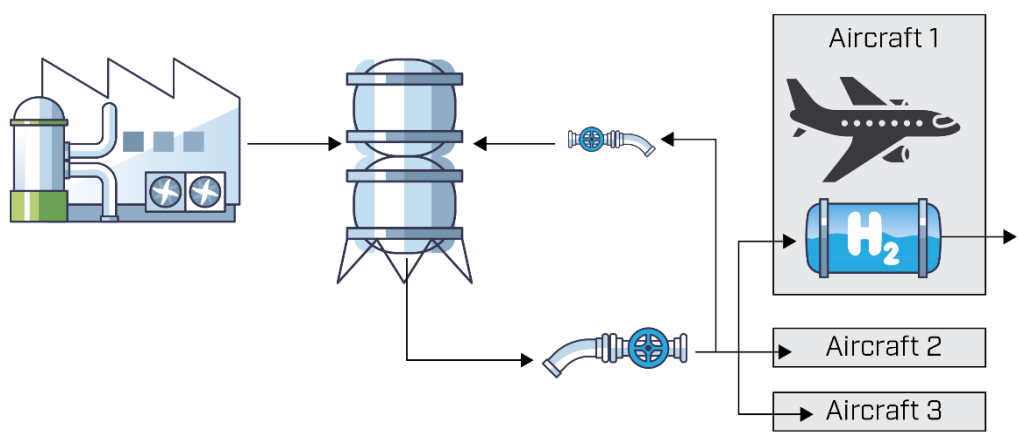

Figure 1: Illustration of the refuelling concept for airports consisting of a transport pipeline from a nearby liquefaction plant and an LH$_2$ distribution pipeline with recycle possibilities.

Highlights

- Simple method to estimate temporal LH$_2$ and CGH$_2$ demand at airports
- Most flight distances require less fuel than the design mission
- Insight into effect of key properties and process parameters for LH$_2$ pipeline systems in airports
- Estimated boil-off can be used by ground support equipment
- Suggestions for operative strategies

# 1   Introduction

Realization of hydrogen as an energy carrier is a major pillar in the energy transition for the transport sector. The direct use of hydrogen as a fuel has reached high technological maturity within road [1], [2] and maritime transport [3], [4]. Several roadmaps for net zero emission aviation points to hydrogen aircraft as a part of the solution [5]. Technology development is ongoing, but delayed due to slow onboard technology maturation and the need for a comprehensive hydrogen ecosystem [6]. Necessary ground infrastructure for hydrogen powered aviation includes airport infrastructures [7] as well as the establishment of production capacity and supply chains for clean hydrogen [8]. Airport infrastructure needs special attention as it is unique in terms of the large potential hydrogen consumption, and the use of hydrogen in liquefied form ($LH_2$) [9], [10].

Today, aircraft are refuelled with Jet-A1 either from pipeline-hydrant systems or fuel bowser trucks. Relatively short $LH_2$ pipeline systems are used at the NASA launch site, but more commonly, $LH_2$ is refuelled from a truck with an intermediate station. Hoelzen et al. [11] studied the economic side of these system concepts, and found that for larger airports, pipeline systems are more reasonable than refuelling by truck bowsers. Brewer et al [12] also favoured pipelines for very large $LH_2$ volumes and transport distances below 64 km. This means $LH_2$ pipelines are relevant both within the airport and to the airport forecourt from a nearby central liquefaction facility.

NASA have reported a high evaporation of $LH_2$ from the $LH_2$ value chain for the space programme[12], [13]. However, the infrastructure at a launch pad operates very differently from what would be the operations at the airport – i. e. every-day use of $LH_2$ and at larger volumes, rather than the more seasonal and low-frequent character of the launch pad. This difference influences the conceptual design and operation of the infrastructure. There is therefore a need for insight into how pipeline systems can be controlled and perform, when used in aviation with large-scale use of $LH_2$ under the strongly time-varying conditions.

In this work, to capture the effect of the time-varying refuelling operations, dynamic process models for an airport pipeline concept are developed. The models are based on the Dymola's TIL library [14] and are used to analyse the effect of system parameters on $LH_2$ distribution to and within the airport. For a given scenario for $LH_2$ refuelling needs at Schiphol Airport (AMS), the amount of boil-off gas (BOG) is estimated. The potential use of this gas by ground support equipment is discussed.

# 2   Materials and methods
## 2.1   Scenario-based estimation of hydrogen demand

Estimates of future large-scale needs for fuelling aircraft ($LH_2$) and ground vehicles (gaseous hydrogen, $GH_2$) at airports are needed for the development of refuelling infrastructure. A time-resolved analysis of future $LH_2$ needs has been conducted for different scenarios of $LH_2$ aircraft market shares. There are large uncertainties related to the market penetration rate of $LH_2$-aviation. This study assumes 50% of single-aisle (200 PAX and up to 2000 nm) and 60% of regional aircraft (100 PAX and up to 500 nm) to be hydrogen-powered by 2050, based on the base case scenario by Hoelzen et al. [11]. This is equivalent to the medium scenario used by Van Dijk et al. [10].

With basis in a flight plan for a given airport on a given day, the meridian arc distance to each destination is calculated. This is the most efficient route, but in reality the additional fuel consumption for routing is estimated to be 12% [15]. The Jet-A1 fuel consumption for a given distance is then calculated using the Base of Aircraft Data (BADA). Flight profiles and fuel needs for every aircraft model and corresponding flights are calculated. The method was validated by comparing the calculated fuel demand throughout 2019 for departures from Rotterdam airport with reported consumption. This analysis showed a good agreement, with only a 6% error between the calculated and reported data [10].

To convert fuel consumption from Jet-A1 and conventional aircraft, to prospective LH$_2$-powered aircraft, the respective heating values are used ($LHV_{JET} = 42.80 \frac{MJ}{kg}, LHV_{LH2} = 119.93 \frac{MJ}{kg}$), and a penalty factor is added for accounting for extra drag and potentially heavier systems [16]. Early phase studies of concepts for hydrogen-powered aircrafts indicates an increased onboard energy consumption of 10-20% [17], but there are significant uncertainties and variations. A feasibility study [16] differentiates aircraft types and finds a reduced energy consumption for commuter (-10 %), regional (-8 %) and short range aircraft (-4 %), and increased energy consumption for medium-range (+22 %) and long-range (+42 %). In this study we only consider flights up to 2000 nm as eligible for LH$_2$, i.e. long-range are not considered relevant for LH$_2$ at this point. As a simplification, we add +10% energy consumption for the conversion of an aircraft to LH$_2$-propulsion.

Hydrogen and fuel cell technologies are also discussed as potential energy sources for Ground Support Equipment (GSE) to reduce emissions at airports [18]. Thus, also hydrogen demand for GSE operations is estimated in this study. Testa et. al [18] give an estimate of the H$_2$ demand for every ground handling operation, dependent on the aircraft class. In the presented work, the aircraft class for every flight is identified based on BADA, and the GSE H$_2$ demand for every flight is calculated throughout a day for different scenario. In the high scenario, all ground operations as defined by Testa et. al are powered by hydrogen. In a medium scenario, aircraft pushback and towing and ground power units operate on hydrogen. These equipment categories were identified by KES BV (Part of TCR Group) in the TULIPS project as candidates for hydrogen. In a low scenario, even less GSE will be powered by H$_2$.

### 2.2 Dynamic model

As pointed out by techno-economic studies [11] [12] and the proposed blueprint for hydrogen infrastructure [7], both a medium and high demand scenario for LH$_2$ motivates pipeline systems. For further analysis of airport LH$_2$ infrastructure, we consider the medium demand scenario and pipeline systems.

#### 2.2.1 Transport pipeline from liquefaction plant to fuel farm

The considered system with boundary conditions is visualized in Figure 2. Cryogenic systems are operated in a subcooled state for several reasons, e.g. to avoid flow instabilities and pump cavitation. To avoid vaporization of LH$_2$ throughout the distribution chain, we set as a critical requirement that the LH$_2$ is subcooled when reaching the airport fuel farm, despite heat ingress and pump inefficiencies that add heat to the fluid. The pressure $p$ of 1.1 bara and temperature $T$ of 19.5 K at the liquefaction plant correspond to a subcooling of 1.05 K. The pressure at the airport's fuel farm is also set to be 1.1 bara.

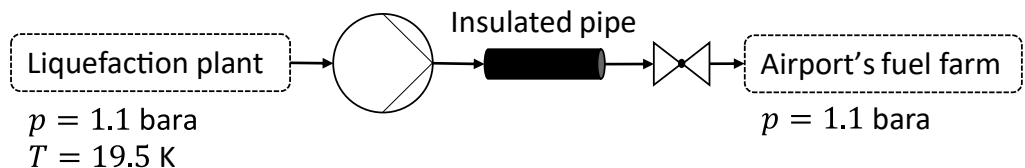

Figure 2: The LH$_2$ transport system from the liquefaction plant to the airport's fuel farm consists of a pump, pipeline and throttling valve.

We consider two cases for transporting LH$_2$ to AMS. Firstly, a 4 km pipeline from a liquefaction plant close to the European hydrogen backbone. Here, we conducted an uncertainty and global sensitivity analysis to study how the subcooling at the fuel farm varies with respect to insulation thickness, pump efficiency, pump sizing and ambient temperature. Secondly, we consider a 25 km long pipeline which corresponds to an import terminal in the Port of Amsterdam. Here, we investigated only the effect of insulation thickness on subcooling at the fuel farm, motivated by the results of the 4 km case.

There are many choices of insulation available for commercial LH$_2$-pipelines [19], [20], with varying cost and insulation properties. Here, a fibre-glass insulation modelled by Hofmann [21] was chosen due to the relatively high thermal performance and availability of data.

We selected a centrifugal pump. The pump curves are defined by the parameters $\eta^{pump}, dp_0^{pump}$ and $V_0^{pump}$, which are the efficiency at the best operating point, the pressure increase for zero flow rate and the volumetric flow rate when the pressure increase is zero, respectively. The pump is described in more detail in the Supporting Information.

For the case of a 4 km pipeline, we consider that a suitable pipeline diameter is either 0.16 m (6'') or 0.22 m (8''). The 6'' pipe causes a higher pressure drop than the 8'' pipe, calling for a more powerful pump. To compare the 6'' and 8'' case, we fixed one sizing parameter of the pump such that the pressure drop over the valve in Figure 2 was small - between 0.2 to 0.45 bar in all simulations. Specifically, we used a fixed-speed pump with $dp_0^{pump} = 1.35$ bar for the 6'' pipe and 0.6 bar for the 8'' pipe, while Table 1 specifies the range of values for $V_0^{pump}$ used in our uncertainty and sensitivity analysis for both the 6" and 8" case. Although a pressure drop of 0.2 to 0.45 bar over the valve may seem small, flow control is feasible by installing a variable speed drive (VSD) on the pump and use a split-range controller [22]. Then, the pump runs on minimum speed and the valve controls the flow rate for small capacities, while the valve is fully open and the pump's speed controls the flow rate for higher capacities.

The uncertainty and global sensitivity analysis used the uncertain parameters in Table 1. We use SALib's implementation [23] of Sobol's method for computing the sensitivity indices of each uncertain parameter with respect to the output variable of interest (i.e. the subcooling at the airport) [24], [25], [26]. The quasi-random samples generated by Sobol's method are also used in the uncertainty analysis to generate histograms of output variables of interest (e.g. the subcooling at the airport).

Table 1: Uncertain parameters in the uncertainty and sensitivity analysis for the 4 km pipeline. We assume uniform, uncorrelated distributions.

| Uncertain parameter | Low bound | High bound |
| --- | --- | --- |
| Ambient temperature, $T_{amb}$ | 278 K | 308 K |
| Pump's nominal efficiency, $\eta^{pump}$ | 50% | 70% |
| Pump's V0, $V_0^{pump}$ | 0.08 $m^3/s$ | 0.10 $m^3/s$ |
| Insulation thickness | 2.5 cm | 5 cm |
| Pressure drop pump $dp_0^{pump}$, 6" pipeline | 1.35 bar | |
| Pressure drop pump $dp_0^{pump}$, 8" pipeline | 0.6 bar | |

The bounds for the uncertain parameters in Table 1 are selected for the following reasons. For pipe diameters of 6'' and 8'', the insulation thickness in commercial products are approximately 2.5 cm [19]. This product sheet uses, however, a more effective insulation (multi-layer insulation), and therefore we assume that doubling the thickness is reasonable for the high bound. For pumps, we selected a typical efficiency range [27, p. 261]. Values for the sizing parameter $V_0^{pump}$ is motivated in the supplementary information.

For the case of a 25 km pipeline, we investigated the effect of insulation thickness on subcooling at the fuel farm. The simulations used 298 K ambient temperature, and we only consider an 8'' pipe diameter. In the analysis we set the pump efficiency to 60%, $V_0^{pump} = 0.11 \ m^3/s$ and $dp_0^{pump} = 1.35$ bar.

### 2.2.2 Aircraft and fuel farm tanks

The fuel farm tank and aircraft tanks are modelled as separators in the TIL library, with the assumption that liquid and gas phase are in thermodynamic equilibrium. During refuelling of the aircraft tank, we assume LH$_2$ is spray-injected from the top, leading to rapid equilibration.

The literature is scatted on the required LH$_2$ tank capacity for SMR aircrafts. Some studies indicate a tank volume around 6000 kg and larger, and boil-off rate (BOR) around 5 % [28], [29], [30]. These studies assume mission profiles associated with longer flights than those identified in our LH$_2$ need scenario and can be seen as conservatively sized tanks. As shown in the LH$_2$ need scenario (Figure 6), most departures require less than 1000 kg of fuel. Therefore, refuelling of a small (600 kg) and large (6200 kg) tank will be analysed.

In our model, the large aircraft tank is set to 96 m$^3$ including 3 % ullage and 5 % as minimum fill level, giving a top-up capacity of 6200 kg. The heat ingress is set to 1.3 kW which corresponds to a BOR of 5.2%. Aircraft tanks have a higher BOR than storage tanks since there are other important constraints for tank design, e.g. a suitable geometry and a low weight. The maximum operation pressure (MOP) is set to 1.7 bara following Gomez [29].

The small tank is set to 9.3 m$^3$, with a top-up capacity of 600 kg LH$_2$. The fill level and ullage are similar to the large tank. The BOR is of 5.3 % (0.13 kW) and MOP of 1.7 bara.

The largest LH$_2$ storage tanks with available information on BOR is 3450 m$^3$ (including ullage), with a BOR of 0.075 % [31]. Fuel farm tanks that continuously supply LH$_2$ to aircraft may have a shorter holding time than space centre tanks and thus can save cost by reducing insulation. In our model, the fuel farm storage tank is therefore set to 8000 m$^3$, which gives a holding time of 1.7 days for the AMS 2050 medium scenario, with a total heat ingress set to 5 kW (BOR 0.17%/day).

### 2.2.3 Distribution system at the airport and operating principles

We consider that aircraft are refuelled between 06:00 and 22:00 (16 h), with no filling at night. However, the system must stay at LH$_2$ temperature overnight to avoid cooling delays in the morning. Each aircraft receives 6200 kg of LH$_2$ in discrete refuelling events, and there are 58 aircraft to fill. Thus, we consider a busy day at AMS where 360 tons is refuelled, which is slightly more than the 330 tons per day received from the liquefaction plant.

Aircraft are refuelled within minutes, and the refuelling time must be kept short to minimize ground time and thereby improve fleet utilization. We have set the refuelling rate to 20 kg/s, which is the physically limited flow rate identified by Mangold [32]. This is faster than the refuelling possible with Jet-A1 [33], but the needed pump technology is missing.

To distribute and refuel aircraft with LH$_2$ we consider the system concept shown in Figure 3. A liquefaction plant supplies a steady flow of 330 tons subcooled LH$_2$ per day to the fuel farm. A fixed-speed centrifugal pump distributes LH$_2$ from the fuel farm to the aircraft through a pipeline system.

Each aircraft tank has a separate flow controller (FC1.1 for aircraft 1, FC1.2 for aircraft 2 etc.) where the valve opening is the manipulated variable. The correct quantity of $LH_2$ is transferred for each tank by adjusting the set-points for the flow controllers between zero and the maximum rate of 20 kg/s. A recycle pipeline is included to allow for continuous cycling of $LH_2$ to maintain the $LH_2$-temperature of the system. The recycle line is also required to prevent turning on and off the pump between the filling of each aircraft, which is often desired to reduce wear and tear of the equipment.

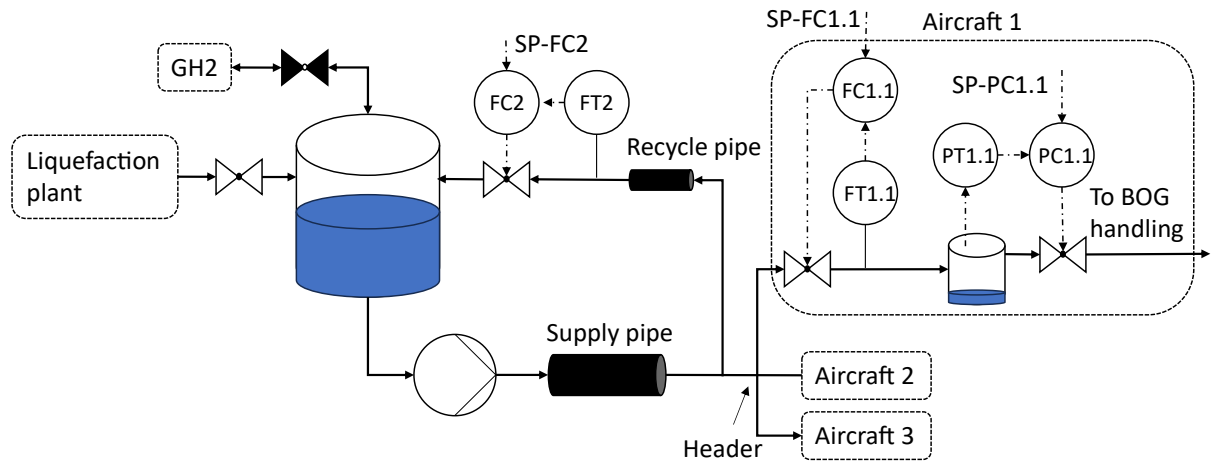

Figure 3: Simplified overview of the delivery system from the fuel farm to the aircraft. FT are flow transmitters (measurements), FC flow controllers and SP-FC its set-point. PT, SP-PC and PC are similar but for pressure control. The pipeline and flexible hose connecting the header to the aircraft is not shown for simplicity

$GH_2$ must be handled at both the fuel farm storage tank and aircraft tank. In this work, the pressure in the fuel farm's storage tank was always in the range 1.03 to 1.3 bara (shown in the supporting information). This we considered to be acceptable, and the $GH_2$ valve is marked as closed (black). For the aircraft tanks, the pressure controller releases $GH_2$ (BOG) to the BOG handling system of the airport to control the tank pressure less than the maximum operating pressure.

The pipelines are modelled with the parameters in Table 2. We assume that the fuel farm is located 2 km from the aircraft, hence, both the supply line the recycle line are 2 km long. In comparison, the ground distance between the storage tank and the launch pad for NASA's space shuttle was 457 m [34]. An 8" flexible hose attaches the aircraft to the main distribution pipeline through a valve. To simplify the simulation, we assumed a constant heat ingress instead of detailed modelling of the insulation. The heat ingress values in Table 2 are motivated in the Supporting Information, and the pump's parameters are also given there. Note that the relatively small diameter of the recycle line (6'') reduces CAPEX, but limits the recycle flow rate of $LH_2$. This recycle flow rate is important to keep the supply- and recycle pipe cold through the night, as we will investigate in the next Section.

Table 2: Specifications of pipelines.

| Pipeline | Length [m] | Diameter [m] | Heat ingress [W/m] |
|---|---|---|---|
| **Supply** | 2000 | 0.45 (18'') | 10 |
| **Recycle** | 2000 | 0.17 (6'') | 3.6 |
| **Header, to hose** | 5 | 0.2 (8'') | 4.6 |
| **Flexible hose from loop to AC** | 10 | 0.2 (8'') | 5 |

A key question is how much hydrogen vaporizes during refuelling (BOG). Refuelling an aircraft tank happens in the timescales of minutes and seconds, but as demonstrated by this study, the refuelling dynamics is dependent on the long-term operation of the airport (timescale of hours and days). Hence, we created two models, one for each timescale. This allows us to focus on relevant effects in each timescale, and to avoid numerical issues in the simulation.

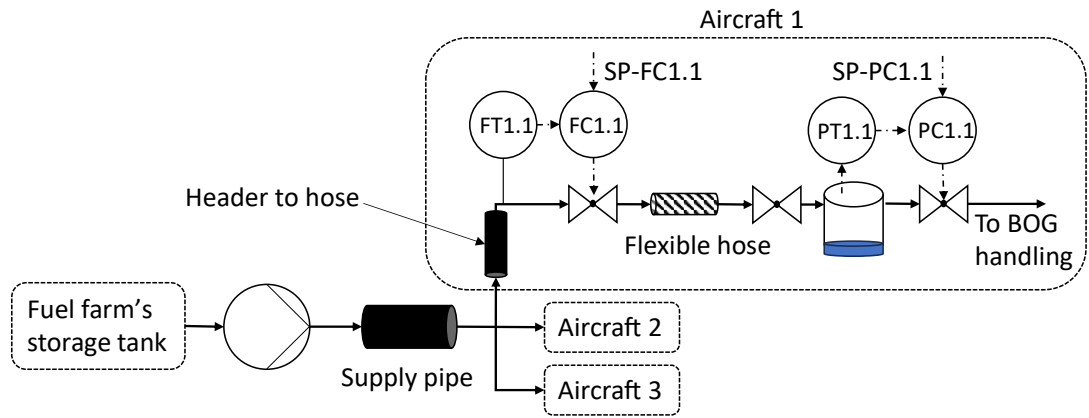

Figure 4: Illustration of the detailed refuelling model created to capture the timescale of seconds/minutes and estimate BOG during refuelling.

Figure 4 shows the detailed model developed for refuelling of a single aircraft tank or several tanks in parallel. The model is used to estimate BOG that must be sent to a BOG handling system. The relatively large supply pipe (2 km, 18") can store heat, and the initial condition of the pipe greatly affects the amount of BOG calculated by our model. To get proper initial conditions for this pipe, which depends on the long-term behaviour of the airport's distribution system, we modelled the system sketched in Figure 5 (timescale of hours/days). In this model, we may study variations in the storage tank in the airport's fuel farm as well as long term temperature variations in the supply and recycle pipe.

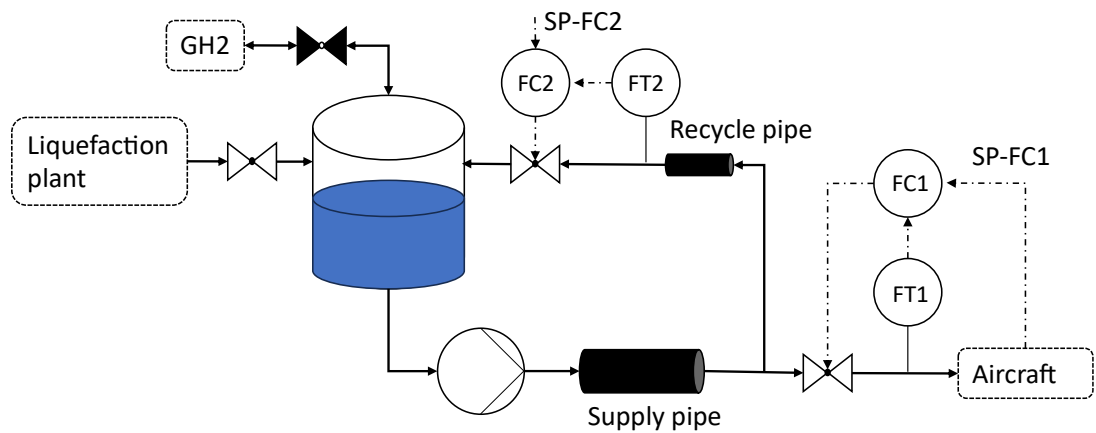

Figure 5: Schematic of the model capturing the behaviour of the airport in the timescale of hours/days, allowing the definition of boundary conditions for BOG calculations. The airport's fuel farm is depicted to the left.

## 3 Results

### 3.1 Estimate of $LH_2$ and $CGH_2$ demand

Figure 6 shows the distribution of flight distances from AMS on an arbitrary day. 42 % of the departures have flight distances below 500 nm (926 km), and 22 % exceed 2000 nm (3704 km). Using the scenario for $LH_2$ aircraft market penetration, the corresponding histogram of $LH_2$ demand per flight is shown in the right panel. 81% of all flights from AMS would require less than 5 t of $LH_2$, 69% than 2 t of $LH_2$, and 15 % less than 600 kg.

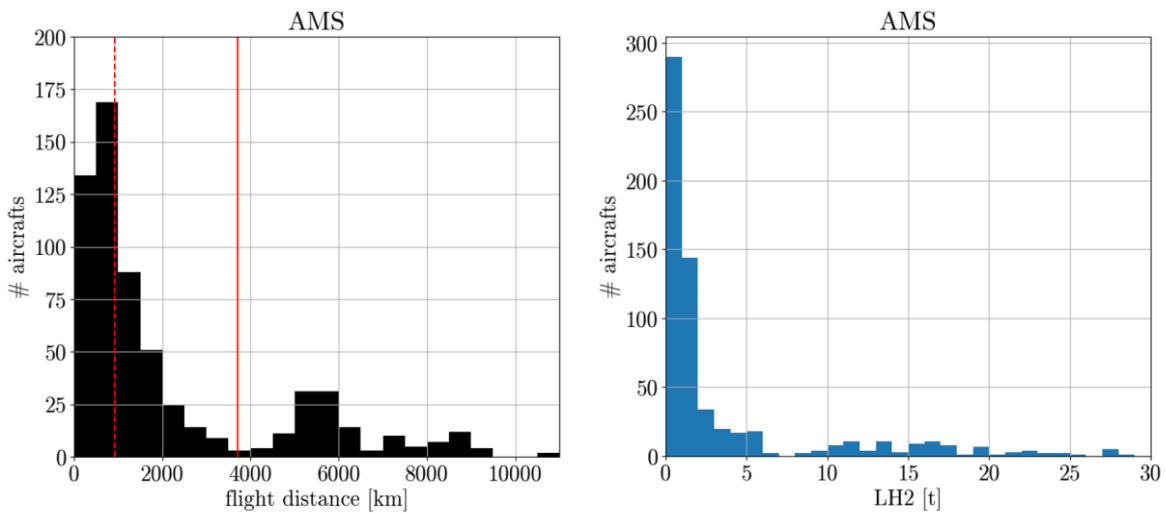

Figure 6: Left: Distribution of flight distances from AMS. The red lines show 500 nm (926 km) and 2000 nm (3704 km) respectively, which is the distance covered by the two hydrogen-powered aircraft available in our scenario. Right: Distribution of required $LH_2$ per departure from AMS.

The time-resolved $LH_2$ demand based on a flight schedule is shown in Figure 7. From 0:00 to 5:00 there are very few departures and thus also little $LH_2$ demand. A first demand peak occurs at 7:00, which corresponds to the peak demand of ~31 t/hour, after which the demand varies between 10 and 30 t/h thoughout the day.

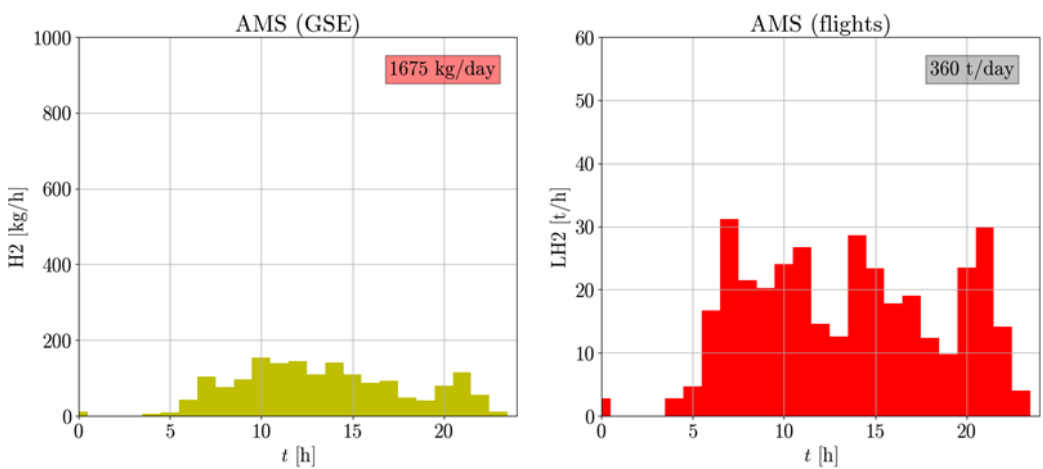

Figure 7: Example of hourly $GH_2$ (left, $H_2$-GSE medium scenario) and $LH_2$ (aircraft, medium scenario) demand throughout a day at AMS.

The medium scenario for hydrogen demand to power GSE throughout a day at AMS is visualized in Figure 7, left panel. Between the medium and high scenario is approximately a factor of five, which shows that it is critical to define which ground handling operations should be powered by hydrogen.

### 3.2 Analysis of pipeline transport from liquefaction plant to LH2 fuel farm

#### 3.2.1 Uncertainty and global sensitivity to pump and insulation parameters for 4 km pipeline

Figure 8 presents histograms showing the uncertainty for selected output variables for the simulation of the 4 km pipeline. The histograms are generated by simulating the model described in Section 2.2.1 with samples from the uncertain input parameters given in Table 1. With the applied properties and within the analysed range for ambient temperature, pipeline insulation and pump efficiency, the subcooling criterion at the fuel farm is met. That is, the temperature is below the saturation temperature of the storage tank, thus indicating a subcooling $T_{SC}^{fuel\,farm}$ larger than 0 K. The 8" pipe has a larger subcooling, i.e. less energy is transferred to the fluid.

However, we see in Figure 8 that the average heat ingress through the pipe, $Q_{ave}^{pipe}$, is smaller for the 6'' pipe than the 8'' pipe. This is due to less surface area of the pipe. Conversely, pump inefficiencies are more important in the 6'' case, which is expected since a small pipe gives a higher pressure drop and requires a more powerful pump.

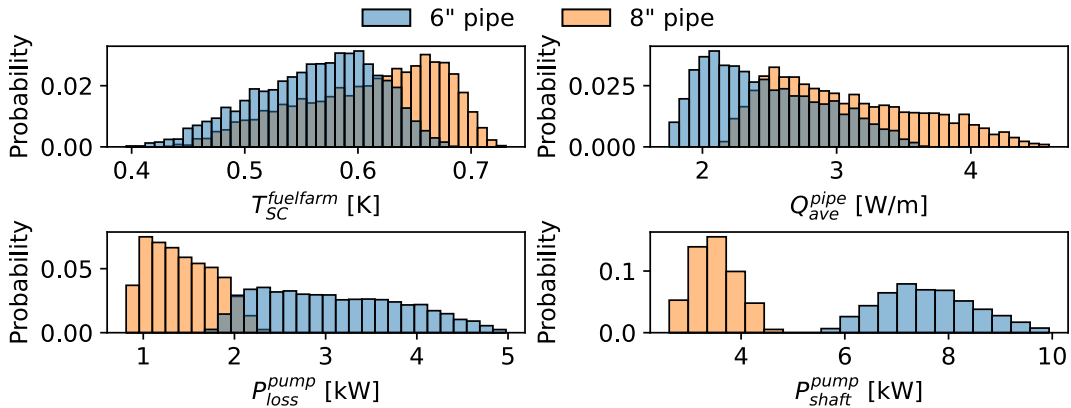

Figure 8: Uncertainty in selected variables: $T_{SC}^{fuel\,farm}$ subcooling in the fuel farm. $Q_{ave}^{pipe}$ average heat ingress into the pipe. $P_{loss}^{pump}$ loss in the pump. $P_{shaft}^{pump}$ pump shaft power. Grey bars are the overlap between 6" and 8" case.

Altogether, we see from Figure 8 that the heat ingress' in the pipeline ($Q_{total}^{pipe} = Q_{ave}^{pipe} \times 4000m$) and from pump inefficiencies $P_{loss}^{pump}$ are substantial for the 4 km pipeline. To understand the underlying contributors to heat ingress, the total sensitivity indices are shown in Figure 9. The uncertainty in insulation thickness is the key driver for the observed variance in the subcooling at the fuel farm, $T_{SC}^{fuel\,farm}$. Effects originating from the pump plays a lesser role, but the pump is more important for the smaller pipe. The performance of the pump is more influenced by uncertainty in the efficiency $\eta^{pump}$ than the sizing parameter $V_0^{pump}$. Shortening of the pipeline has several advantages, as $Q_{total}^{pipe}$ will be smaller, and a lower differential pressure can allow for reducing pump work' and therefore less energy is transferred to the fluid.

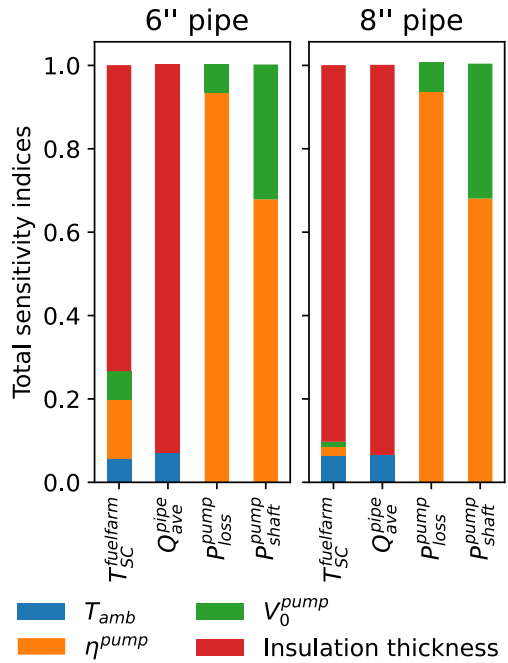

Figure 9: Total sensitivity indices. $T_{amb}$ ambient temperature. $\eta_{pump}$ pump efficiency. $V_o^{pump}$ pump sizing parameter.

### 3.2.2 Requirements to insulation thickness for a 25 km pipeline

The subcooling temperature at the fuel farm and heat ingress through a 25 km pipeline is simulated for various insulation thicknesses as shown in Figure 10. It is observed that to obtain a marginal subcooling at the fuel farm, the insulation thickness must be at least 12 cm, since a thinner insulation yields two-phase flows in our simulations. In practice, one would design the pipe with a margin on the insulation thickness, meaning that the insulation thickness would be more than 12 cm. The increased insulation thickness will increase the cost of the pipeline per meter, and therefore CAPEX of a pipeline for LH$_2$ does not scale linearly with its length.

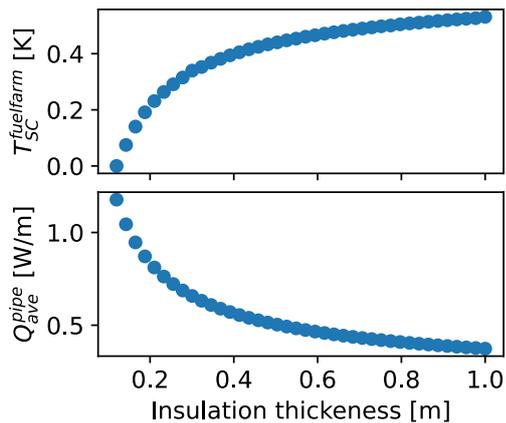

Figure 10: Subcooling in a 25 km pipeline as a function of insulation thicknesses.

## 3.3 Analysis of LH$_2$ delivery system to aircraft and refuelling

### 3.3.1 Behaviour of the airport's distribution system over 40 hours

The long-term behaviour of selected variables in the airport's distribution system is shown in Figure 11 for two daytime periods (16 h each) separated by one night (8 h, grey area). As shown in the top pane, LH$_2$ is pumped to either refuel aircraft, or sent to the recycle loop when no aircraft are being refuelled. The recycle flow rate is a control variable used to maintain the cryogenic conditions in the system. During daytime the minimum recycling rate is set to 0.2 kg/s when aircraft are refuelled, and increased to 3 kg/s when no refuelling occur. At night, the recycle flow rate is set to a level which maintains the cryogenic conditions.

Results from the dynamic simulation shows, as expected, that the pipe is warmest (least subcooling) in the morning, since there has been a long period with low LH$_2$ flow through the pipe. A nightly recirculation rate of 3.8 kg/s gives a colder morning pipe than a recirculation of 2.8 kg/s. However, after some hours into day 2, the subcooling is similar for the two cases. We cannot decrease the recirculation rate much below 2.8 kg/s in-between refuelling operations and at night, since we are very close to losing the subcooling in the recycle pipe in the morning, as seen in the zoom inset box in Figure 11.

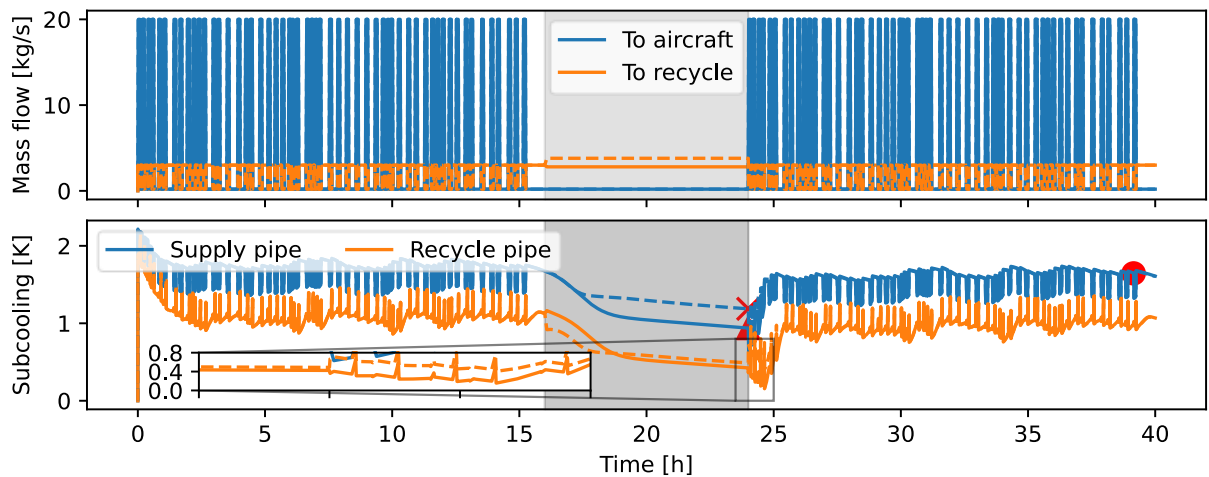

Figure 11: Top pane shows the LH$_2$ mass flows to either refuel aircraft or the recycle. The second pane shows the subcooling at the outlet of the supply and recycle pipe. Grey area marks the night, when no aircraft are refuelled. Solid lines represent medium recirculation (2.8 kg/s), and dashed lines high (3.8 kg/s) recirculation. Red marks indicate refuelling events investigated in our refuelling model.

### 3.3.2 Calculation of hydrogen vapor formation and BOG handling needs

Our hypothesis is that the H$_2$ vaporization (BOG formation) is strongly sensitive to the subcooling condition. To test this, we developed the more detailed refuelling model shown in Figure 4. As initial conditions for temperature and suction pressure (approximated by fuel farm tank pressure, see supporting information), we choose the situations indicated by the red markers in Figure 11. This corresponds to the subcooling state when filling the first flight after a night of high or medium recycle flow, and filling of the last flight at the end of the day.

For all three cases, we consider the following filling regime. First, we fill Aircraft 1 with 20 kg/s. Then, Aircraft 2 and 3 are filled in parallel which means the pump should deliver 40 kg/s. Figure 12 shows the resulting fill level and pressure in the aircraft tanks, the flow of LH$_2$ into the tanks and gaseous hydrogen vented from the tanks and sent to the vapor BOG management system. This system is outside the scope of this work, but can be e.g. the suction of a compressor used to either power ground support equipment or for returning gas to the central LH$_2$ storage in a gas return line.

For each aircraft, the pressure in the aircraft tank increases as the fuelling progresses. When the pressure reaches the maximum operating pressure of 1.7 bara, GH$_2$ is vented from the tank to a separate GH$_2$ system. The amount of gas ejected from the tank varies for the three refuelling events. Less H$_2$ is vented if the recycle rate is high during the night (dotted lines – 3.8 kg/s recycle rate), compared to the medium nightly recycle rate (dashed lines – 2.8 kg/s recycle rate). This is due to the larger degree of initial subcooling. The amount of $GH_2$ ejected decreases quite substantially from filling aircraft 1 to aircraft 2 and 3. There are two reasons for this. First, that the system gets colder when we pump a large amount of $LH_2$ through it for filling aircraft 1. Second, increasing the mass flow through the centrifugal $LH_2$ pump decreases the discharge pressure of the pump, and there are therefore less losses due to throttling when filling aircraft in parallel. Actually, the flow control valves are fully open during parallel refuelling in our simulations (not shown for brevity), and there are therefore no throttling losses in this case. However, this is due to a slightly under-dimensioned pump since we see that the $LH_2$ fuel flow to aircraft 2 and 3 are below the set-point of 20 kg/s (40 kg/s in total) when the aircraft pressures are 1.7 bara.

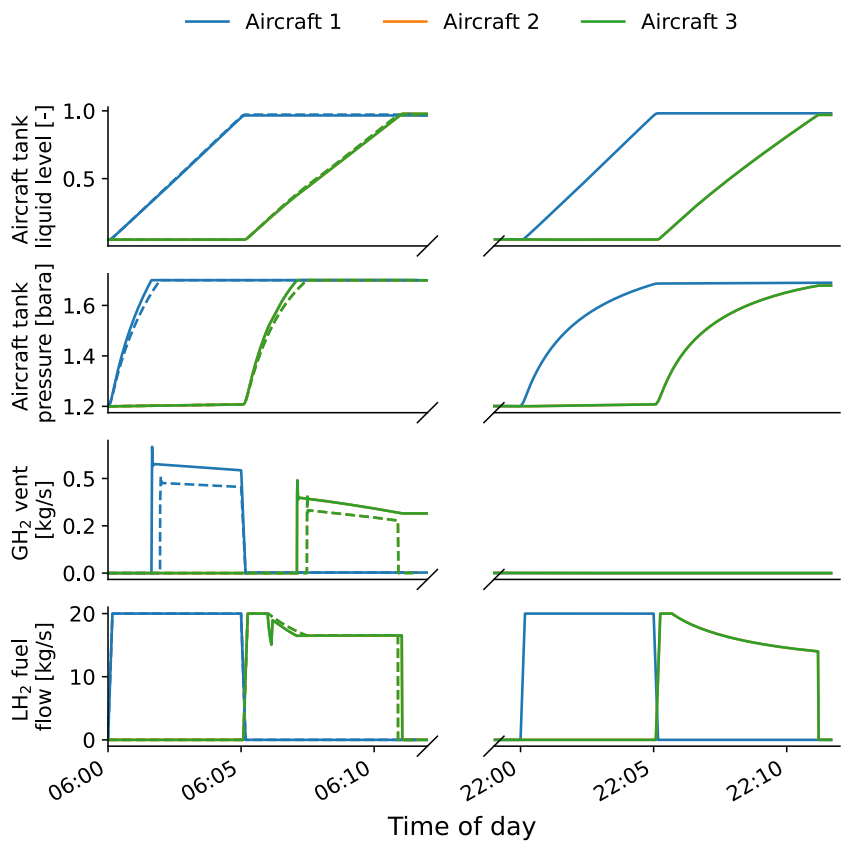

Figure 12: Development in the three aircraft tanks for the filling with a nightly recirculation rate of 2.8 kg/s (solid lines) and 3.8 kg/s (dotted lines). Aircraft 2 and 3 are filled in parallel, and their lines are therefore overlapping.

When we start with the coldest pipes, which is for the aircraft filled at 22:00, the pressure in the tanks reaches 1.68 bara at the end of the filling. This is slightly below the maximum operating pressure of 1.7 bara and venting does not occur. The total amount of $GH_2$ vented during each filling is given in Table 3. To understand the effect of the possibly over-sized tank, refuelling simulations were also performed for a 600 kg tank. The BOG is naturally reduced, but the relative evaporation is higher (2.1 % vs 1.9%), and more refuelling events will be needed to reduce BOG.

Table 3: Accumulated vented $GH_2$ from each aircraft during the fueling operation.

| Refueling case | | | Boil-off gas | | |
|---|---|---|---|---|---|
| Tank size, kg $LH_2$ | Time | Nightly recycle rate | Aircraft 1 | Aircraft 2 | Aircraft 3 |
| 6200 | 06:00 | 2.8 kg/s | 117 kg | 86 kg | 86 kg |
| 600 | 06:00 | 2.8 kg/s | 13 kg | 12 kg | 12 kg |
| 6200 | 06:00 | 3.8 kg/s | 88 kg | 63 kg | 63 kg |
| 600 | 06:00 | 3.8 kg/s | 10 kg | 9 kg | 9 kg |
| 6200 | 22:00 | N.A. | 0 kg | 0 kg | 0 kg |

If the plane is filled at 22:00 and left overnight 77 kg BOG ($GH_2$) per aircraft will be generated due to heat ingress in the tank. A summarized, crude estimate of the total BOG generated for the entire distribution system is provided in Table 4, under the assumptions of this work. It is assumed that 58 aircraft are filled with 6200 kg $LH_2$ each throughout the day. The refuelling BOG low estimate is 71 kg + 2×53 kg + 55×0 kg = 176 kg, while the high estimate is 92 kg + 57×73 kg = 4243 kg. The high scenario for aircraft overnight holding considers 10 aircraft.

Table 4: Crude estimate of BOG generation for the airport distribution system.

| | Low | High |
|---|---|---|
| Refuelling BOG | 176 kg | 4243 kg |
| Stationary storage BOG | 356 kg | 356 kg |
| Aircraft overnight holding BOG 77 kg/ac | 0 kg | 770 kg |
| Total amount of BOG generated | 532 kg | 5369 kg |

A BOG estimate of 5.4 tons/day are in the same ballpark as the scenario for GSE $GH_2$ need (1.7 – 9 tonnes). This means that there is a valuable potential in investigating integration of $CGH_2$-GSE systems with BOG management systems.

Vapor formation in the supply system shall be avoided, but even when maintaining the subcooling criteria it can be hard to avoid vaporisation. The bottom pane in Figure 12 shows a dip in fuel flow at around 6 minutes for the mass flow rate to aircraft 2 and 3 for the case of a medium nightly recirculation rate (2.8 kg/s). This dip in the mass flow rate is due to vapor formation in the large supply pipe. Filling aircraft in parallel means an increased mass flow rate through the pump, which corresponds to a reduced discharge pressure since we assumed a centrifugal pump. Since the boiling point of $LH_2$ depends on temperature and pressure, the sudden decrease in pressure may cause vapor formation (flashing) in the pipeline, and a two phase-flow leading to reduced mass flow. This is shown more detailed in the supporting information. Strategies to avoid this effect can be to reduce the refuelling rate or avoid parallel refuelling.

## 4 Discussion

### 4.1 Aircraft tank

SMR aircraft are a key platform for emission reduction from aviation. Available scientific literature on LH$_2$ tank sizes for SMR aircraft deviate from the estimated LH$_2$ needs for typical operations (Figure 6). The literature typically considers longer flight durations as design missions, which can cause this deviation. To address the uncertainty in aircraft tank size, we analyse the refuelling of both 6200 and 600 kg tanks. The overall refuelling behaviour is similar, but a higher relative and lower absolute formation of H$_2$ BOG vapor is seen (Table 3).

An important design parameter for the aircraft tank is the maximum operating pressure. A higher operating pressure allows for more vapor formation and less venting, but this influences the refuelling system as discussed below.

### 4.2 Subcooling criteria and pipeline transport to the fuel farm

We consider two pumps in this paper. First, from the liquefier to the airport's fuel farm, and then from the fuel farm to the aircraft. Importantly, there is no intermediate cooling between the pumps. Therefore, we may operate the liquefier to get more subcooling on the suction of the first pump, and this also benefits the second pump. Subcooling increases the net positive suction head available (NPSHa), which must be higher than the NPSH required (NPSHr) stated by the pump supplier for safe operation. Therefore, we obtain a minimum subcooling criterion for the system by considering the NPSHa and NPSHr. However, we do not have any values for NPSHr since LH$_2$ pumps in our considered capacity range do not exist yet.

The results of the global sensitivity analysis are subject to assumptions in our model and the ranges of the considered uncertain variables. Importantly, we have designed (modelled) the fixed-speed pump such that the pressure drop over the throttling valve is small. In practice, it is likely that a fixed-speed pump would have a higher delivery pressure as a safety margin to ensure that the correct quantity of LH$_2$ is transported to the airport. Increasing the delivery pressure translates to a more powerful pump, implying that that more energy is transferred to the LH$_2$ and it is likely that the pump's parameters would play a more prominent role in the sensitivity analysis. Therefore, we stress that our analysis assumes a close-to-optimal pump operation, which is feasible by installing a VSD on the pump and apply split-range control as noted in Section 2.2.1. However, a VSD is expensive and may introduce vibration-related issues.

Other important error sources in our model are that we use a generic pump model, and that the applied pressure drop model of Konakov [35] is not verified against experimental LH$_2$ measurements due to lack of data. The pipeline model also assumes a straight pipe, but note that bends and other fittings in a physical pipe can be converted to *equivalent lengths*, which is a common engineering method [36].

### 4.3 Operation of the distribution system

The distribution system consists of a pump, the large storage tank in the fuel farm and the small aircraft tanks. The model assumes thermodynamic vapor-liquid equilibrium (VLE), which may cause overestimation of the BOG formation. Non-equilibrium behaviour in real stationary systems, such as the storage tank, cause stratification of the gas phase as a result of the heat ingress from the ambient warms the gas as well as evaporating the liquid. Furthermore, the pumped LH$_2$ is extracted from the bottom of the tank. The pressure of the liquid column produces a certain degree of subcooling; a 10 m LH$_2$ column gives approximately 0,17 K subcooling. These factors lead to more subcooling of the LH$_2$ at the suction of the pump, and the temperatures in all pipes would be lower.

This implies that the pressure increase in the aircraft's fuel tanks would be slower and less BOG generated.

The maximum operating pressure (MOP) and thermal performance are important design parameters for the aircraft tank. For simplicity, we assumed that all aircraft tanks have the same MOP. Higher tank pressures may be permittable as aircraft tanks are under development in other research programmes, e.g. COCOLIH$_2$T [37]. The design criterion for the refuelling pump is that it can provide the correct fuel flow for the tank with the highest MOP. Increasing the MOP allows storing more GH$_2$ in the aircraft tank before the GH$_2$ is released to the BOG handling system. However, this calls for a more powerful refuelling pump, which will transfer more energy to the fluid and thereby increase BOG formation. This undesirable effect can be minimized by installing a VSD on the pump and apply split-range control for the fuel flow. In this case, the pump will minimize the energy consumption and BOG formation even if there are different MOPs for the aircraft tanks to be refuelled. Investigating the effect of the MOP on the BOG formation with and without VSD on the pump is interesting further work.

Although the uncertainties are large, the results still give valuable insights into the variation of subcooling in the system through a day and the corresponding effect on BOG formation. The nightly recirculation rate is a free variable which can be used to keep the long supply pipe cold and reduce vapor formation in the morning. An "optimal recirculation rate" which balances the power consumption from the pump versus the amount of GH$_2$ formed may exist, and this is an interesting topic for future work.

### 4.4 Use of BOG in GSE operations

BOG (GH$_2$) should not be released to the atmosphere, but it can be compressed and used for ground support equipment. The daily BOG scenarios (0.5 tpd – 5.4 tpd) correspond well with the estimated need for GH$_2$ for ground support equipment (1.7 tpd - 8.9 tpd). Gaseous hydrogen is used by GSE at various levels throughout the day, mirroring the flight schedule. As demonstrated by the refuelling simulations, more BOG is formed in the morning than in the evening – and BOG will therefore not mirror the flight schedule. This causes a temporal mismatch between the formation and use of BOG, which necessitates storage.

Note that although the GH$_2$ is used, the energy used for liquefying H$_2$ is partly lost. The most energy intensive step in the hydrogen value chain is the hydrogen production itself, with about 55 kWh/kg H$_2$ produced through electrolysis [38]. State of art research and industrial processes for larger scale liquefaction consumes 6-10 kWh/kg respectively [39], [40] and thereby add 10 - 17 % to the energy consumption. However, gaseous hydrogen can still have low temperatures and higher density, which may lead to more efficient compression into GSE-vehicles or gaseous storage. Currently compression for vehicles consumes 1.7-6.4 kWh/kg [41]. Finding solutions to fully integrate BOG handling into hydrogen refuelling stations can therefore give significant savings. The residual cold may also be used as a shield gas in the pipelines to reduce BOG generation.

## 5 Conclusion

In summary, the presented analysis shows that

- Continuous recycling of LH$_2$ at a moderate rate is needed to maintain subcooling in the distribution system. The formation of surplus vapor is substantially higher when filling the first plane in the morning (warm pipes) than the last plane in the evening.
- In our analysis of pipeline transport from the liquefaction plant to the fuel farm, the global sensitivity analysis revealed that the insulation thickness is the key uncertain variable to preserve the subcooling. We showed that a longer pipeline requires thicker insulation than a shorter pipeline, which implies that the CAPEX of $LH_2$ pipelines is non-linear with respect to length.
- System BOG generation is very sensitive to pump operation and pressure drops. Filling aircraft in parallel means an increased mass flow rate through the pump, which corresponds to a reduced discharge pressure. The reduction in pressure can cause vapor formation if the pipeline has marginal subcooling. Vapor formation can be avoided by increasing the pump's discharge pressure, which can be achieved by either reducing the refuelling rate or avoiding parallel refuelling.
- Estimated BOG generation of the modelled system corresponds well with the energy demand of GSE equipment at the airport. Other usages of the BOG $GH_2$, such as for emergency power generation, may also be considered if the demand for GSEs should be lower.

## 6  Further work

In general, experimental data is needed for validation of the models. To improve concept design and the understanding of infrastructure for $LH_2$ aircraft, further work on the optimization of the aircraft tank size, maximum operating pressure and insulation is needed. This influences the refuelling system requirements. Some components (the pump) in the proposed system does not exist yet, and these need to be developed. Further optimization of operating principles can be done, to balance the power consumption of the pump, subcooling of the liquefying process and allowable evaporation.  For this, a process concept for integration of cold, evaporated hydrogen must be investigated.

Acknowledgements

This document has been created with or contains elements of Base of Aircraft Data (BADA) Family Release which has been made available by EUROCONTROL to SINTEF Energy Reserach. EUROCONTROL has all relevant rights to BADA. ©2019 The European Organisation for the Safety of Air Navigation (EUROCONTROL). All rights reserved. EUROCONTROL shall not be liable for any direct, indirect, incidental or consequential damages arising out of or in connection with this product or document, including with respect to the use of BADA.

Funding

This work was supported by the TULIPS project (European Union) through Grant Agreement No. 101036996. This article is based D2.5 in this project.

Declaration of competing interest

The authors declare that they have no known competing financial interests or personal relationships that could have appeared to influence the work reported in this paper.

CRediT authorsip contribution statement

**Halvor Aarnes Krog**: Conceptualization, methodology, formal analysis, software, visualization, writing – original draft. **Yannick Joss**: Conceptualization, methodology, formal analysis, software, visualization, writing – original draft. **Hursanay Fyhn**: Validation, Formal analysis, Writing – original draft **Petter Nekså**: Conceptualization, methodology, supervision, funding acquisition, writing – original draft. **Ida Hjorth**: Conceptualization, methodology, formal analysis, investigation, visualization, writing – original draft, funding acquisition

## Appendix A.    Supporting information

### A.1 Pipeline transport from the liquefaction plant to the LH₂ fuel farm

#### A.1.1    Model of the LH₂ pump

The pump was simulated as the 2nd order Pump in TIL's library. The pump is specified through the parameters

- $V_0^{pump}$ [$m^3/s$]: the volume flow when the pressure increase is zero.
- $dp_0^{pump}$ [bar]: the pressure increase at zero volume flow
- Efficiency $\eta^{pump}$ [-]: efficiency of best operating point.

Figure 13 shows the physical meaning of these parameters in the pump chart, for three different values of $V_0^{pump}$. In the presented system, a transport rate of 330 tons/day is specified, which corresponds to approximately 0.054 $m^3/s$ of LH₂, which is indicated by the red dashed line.

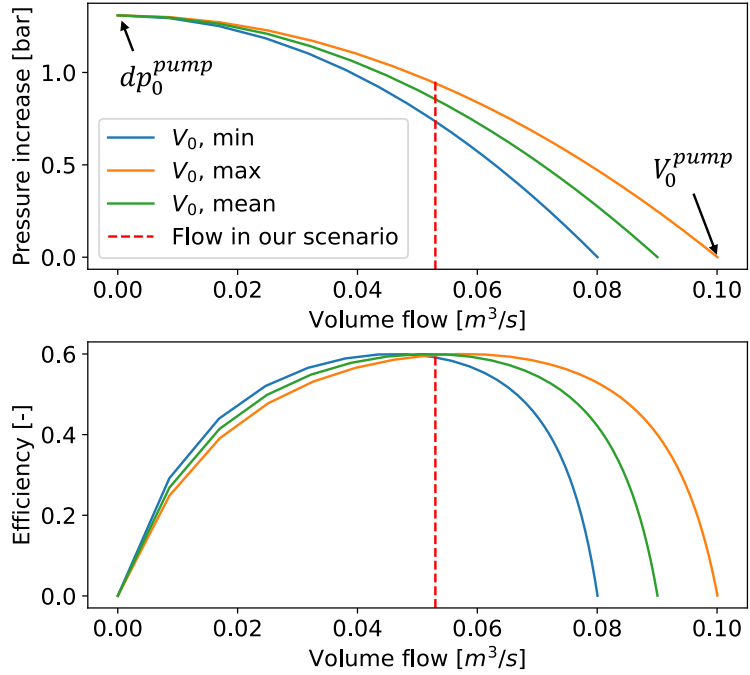

Figure 13: Pump curves with $dp_0^{pump} = 1.35 bar$, efficiency of 60% and values of $V_0^{pump}$ corresponding to 0.08 $m^3/s$, 0.09 $m^3/s$ and 0.1 $m^3/s$ (annotated in the chart), respectively. The flow rate corresponding to 330 tpd LH₂ are shown as the red dashed line.

### A.2 BOR calculation tank

The calculation of heat ingress for the airport fuel farm tank follows the approach from Berstad [42]. We assume a well-insulated, spherical storage tank with ambient temperature of 35 Celsius, i.e. the high bound for ambient temperature in Table 1 (main article), representing a hot summer day. The heat ingress is estimated by selecting an overall heat transfer coefficient $UA = 0.009 \frac{W}{m^2 K}$ and inside the tank the saturation temperature of LH₂ at 1.2 bara (i.e., 20.86K).

To compare the heat ingress of the fuel farm with current build tanks of 3450 m³, we use the reported BOR values[31], from which we derive a heat ingress of 0.0015 W/m²K using a network model tool [43]. The UA value of our 8000 m³ tank corresponds to 0.18 % BOR and is around 6 times larger than our estimate for the 3450 m³ tank.

## A.3 Airport's distribution system

### A.3.1 Heat ingress values for the pipelines and flexible hose.

This Appendix motivates the heat ingress values for Table 2 (main article).

Heat ingress values for the 6'' and 8'' pipes were taken as the high bound from Figure 8. The heat ingress for the 8" flexible hose is set 10% higher than the 8'' pipe. We estimated the heat ingress for the 18'' pipe by doing a linear regression based on tabulated heat ingress (for liquid $N_2$) [19] for pipe sizes ranging from 2'' to 8'', see Figure 14. We used the linear model to extrapolate heat ingress to 18'' and then added 10% to the heat ingress as our fluid is $LH_2$.

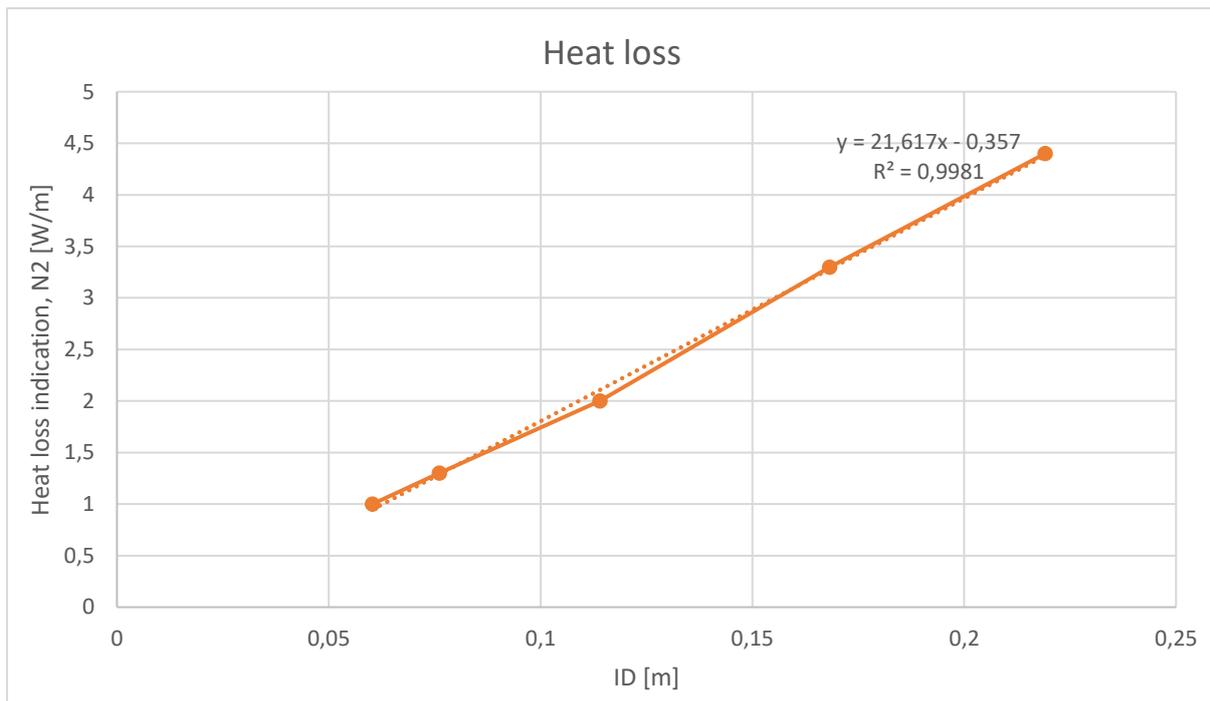

Figure 14: Heat loss from pipes based on $LN_2$.

### A.3.2 Pump parameters

The pump is specified with the parameters $dp_0^{pump} = 0.9$ bar, $V_0^{pump} = 0.8 m^3/s$ and $\eta^{pump} = 60\%$.

### A.3.3 Initial conditions pump suction pressure

The fuel farm tank is continuously being refilled with $LH_2$ supplied by a liquefier at a rate of 330 tpd. The liquid level and pressure in the fuel farm's storage tank are highest in the morning and lowest in the evening, as shown Figure 15. In the BOG calculations a very low pressure drop between the fuel farm tank and the pump is assumed, such that the fuel farm pressure is used as initial conditions for the pump's suction pressure.

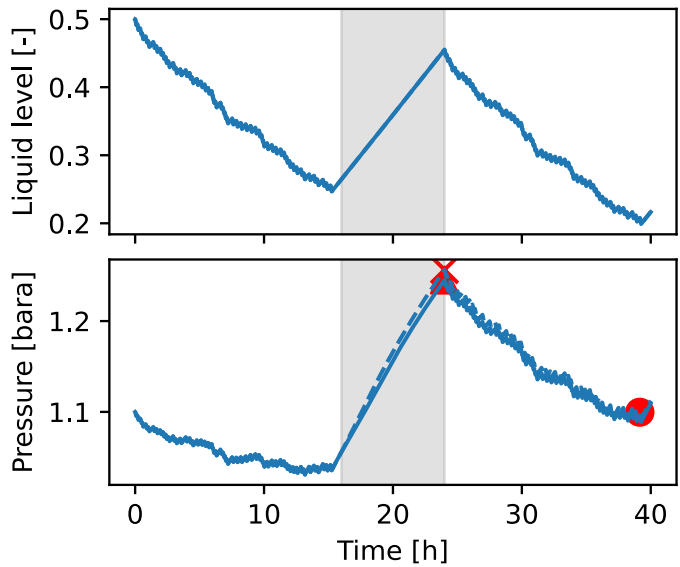

Figure 15: Pressure and liquid level in the fuel farm tank. BOG is calculated for the refueling events indicated with red markers.

### A.4 Flashing at high flow rates

A dip in mass flow rate is observed when aircraft are refuelled in parallel in the morning, after a night with 2.8 kg/s of recirculation. The absolute temperature, subcooling temperature and liquid mass fraction in the main supply pipe is shown in Figure 16. Parallel refuelling of aircraft starts after 5 minutes, after which there is vapor formation in the pipe within the first minute. The resulting mass flow rate dip occurs thereafter (approximately at 6 minutes, see Figure 12 in the main article); the delay is explained by the control valve opening, which is very fast and saturates at around 6 minutes.

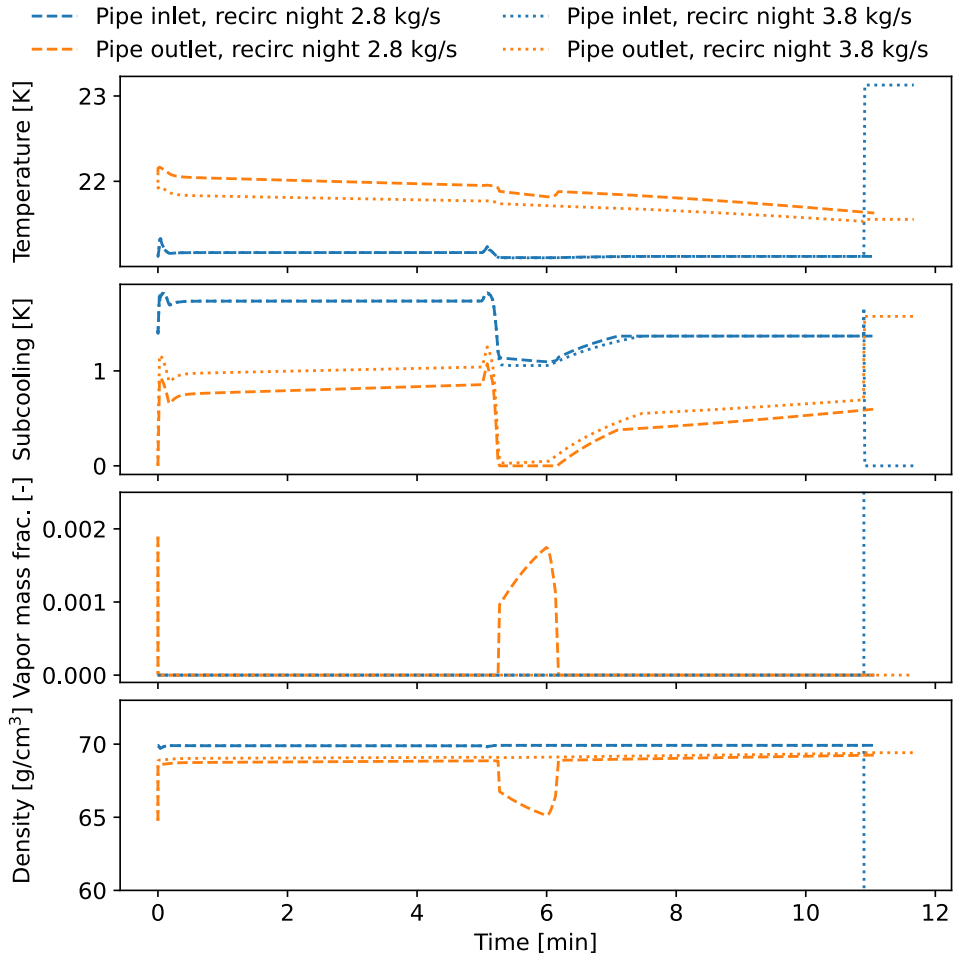

Figure 16: Vapor formation in the supply pipe when refuelling aircraft in parallel.

### A.5 Hydrogen needs for GSE

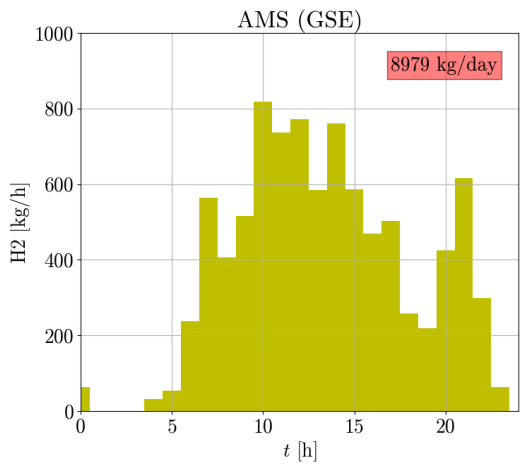

Figure 17: Hydrogen needs for GSE operations, high scenario